\documentstyle [prl,aps,preprint]{revtex}

\begin{document}
\title{Galilean non-invariance of geometric phase}
\author{Erik Sj\"{o}qvist$^{1,2,}$\footnote{E-mail: 
erik.sjoqvist@philosophy.oxford.ac.uk}, Henrik Carlsen$^{3,}$\footnote{E-mail:
henrik.carlsen@kvac.uu.se} and Harvey R.
Brown$^{2,}$\footnote{E-mail: harvey.brown@philosophy.oxford.ac.uk}}
\address{$^{1}$Department of Mathematical Sciences, University of Durham,
Durham DH1 3LE, UK \\
$^{2}$Sub-Faculty of Philosophy, Oxford University, 10 Merton Street,
Oxford OX1 4JJ, UK \\
$^{3}$Department of Quantum Chemistry, Uppsala University, Box 518, S-751 20 
Uppsala, SWEDEN \\}
\maketitle
\begin{abstract}
It is shown that geometric phase in non-relativistic quantum mechanics is not
Galilean invariant.
\end{abstract}
Consider, in the context of non-relativistic quantum mechanics,  
a system undergoing cyclic evolution during the interval $[0,T]$, so that its 
final and initial states coincide up to a global phase: $|\psi (T) \rangle =
e^{i\phi} |\psi (0) \rangle$, with $\phi$ being an arbitrary real number.
Such evolution defines a closed curve in projective Hilbert space ${\cal P}$
(the space of rays in the Hilbert space ${\cal H}$ of the system). Following
the work of Aharonov and Anandan \cite{aharonov}, itself a generalisation of
the seminal Berry \cite{berry} analysis of particular systems undergoing
adiabatic evolution, it is known that the phase $\phi$ can be decomposed
into a geometric and dynamic part; the geometric part, denoted here by 
$\gamma^{AA}$, determined by removing the accumulation of local phase
changes\footnote{The local phase change 
$\delta \eta (\psi_{t} ,\psi_{t+\delta t} )$ is defined as the phase difference
between two infinitesimal close state vectors 
$|\psi (t)\rangle$ and $|\psi (t+\delta t)\rangle$, i.e. 
$i\delta \eta (\psi_{t} ,\psi_{t+\delta t} )
= (\ln \langle \psi (t)|\psi (t+\delta t)\rangle -\ln \langle 
\psi (t+\delta t)|\psi (t)\rangle )/2 \approx 
\langle \psi (t)|d/dt|\psi (t) \rangle \delta t$.} 
from the global phase $\phi$, i.e.
\begin{equation}
\exp (i\gamma^{AA} [\psi ]) = \langle \psi (0) |\psi (T) \rangle \exp \left( 
-\int_{0}^{T} \langle \psi(t) |\frac{d}{dt}|\psi (t)\rangle dt \right) ,
\label{eq:AA}
\end{equation}
where $\gamma^{AA} [\cdot ]$ is a functional of the cyclic path 
$|\psi (t)\rangle$ in ${\cal H}$. The Schr\"{o}dinger equation and 
(\ref{eq:AA}) make it clear that the dynamic phase $\gamma_{d}$ is given by
\begin{equation}
\gamma_{d} = -i\int_{0}^{T} \langle \psi (t) |\frac{d}{dt}|\psi (t)\rangle dt
= -\frac{1}{\hbar} \int_{0}^{T} \langle \psi (t) |H| \psi (t) \rangle dt .
\end{equation}
Here the operator $H$ is the Hamiltonian generating the evolution of the
system in the interval $[0,T]$.

Now $\gamma^{AA}$ is reparametrisation invariant, i.e. independent of the
speed at which the path is traversed.
Furthermore, it is projective-geometric in nature. Given a closed curve in
${\cal P}$, there is an infinity of Hamiltonians generating motions in 
${\cal H}$ which project onto the curve. The phase $\gamma^{AA}$ is indifferent
to the choice of Hamiltonian, and depends only on the curve in ${\cal P}$.
In the light of these properties, the geometric phase can be interpreted as the
anholonomy transformation associated with a natural background connection
(curvature) in that space\footnote{A recent resource letter on geometric 
phases is found in Anandan {\it et al.} \cite{anandan1}}.

It was pointed out by Anandan \cite{anandan2} that the closure property of
a curve in ${\cal P}$ is frame-dependent. To see this, note that the state 
of the system relative to the frame moving with velocity ${\bf v}$ relative
to the laboratory frame, $|\tilde{\psi} (\tilde{t}) \rangle$ $(\tilde{t}=t)$, 
is obtained from the state defined relative to the latter frame by the action 
of a unitary operator (passive Galilean boost) $U_G$: 
$|\tilde{\psi}(\tilde{t})\rangle=U_{G} (t)|\psi(t)\rangle$, the form of 
$U_G$ given by\footnote{See, e.g., Peres \S 8.8 
\cite{peres}, and particularly Fonda and Ghirardi \S 2.5 \cite{fonda}. 
These discussions extend to the case of a particle moving in an external 
scalar potential; the more general case involving an additional vector 
potential, in which (\ref{eq:passive}) below is still valid, is discussed in 
Brown and Holland \cite{brown}.}
\begin{equation}
U_{G}(t) = e^{i{\bf v}\cdot (-m{\bf Q}+t{\bf P})/\hbar } =
e^{-im{\bf v}\cdot {\bf Q}/\hbar}e^{i({\bf v}\cdot {\bf P}
-m{\bf v}^{2}/2)t/\hbar}. 
\label{eq:passive}
\end{equation}
Here, $\bf{Q}$ is the position operator, $\bf{P}$ the canonical momentum 
operator, $m$ the mass of the system and for the last equality in
(\ref{eq:passive}) we used the operator identity $e^{A+B}=e^Ae^Be^{-[A,B]/2}$
valid for operators $A$ and $B$ which commute with their commutator. 
It is clear, given the non-trivial time dependence of $U_G$, that whether the 
evolution of the system in the interval $[0,T]$ is cyclic depends on the state 
of motion of the observer.

It follows from this observation that the very condition required for the 
definition of the Aharonov-Anandan geometric phase $\gamma^{AA}$ can be met 
relative to at most one inertial frame.
Indeed, recognition that the closure property of curves
in ${\cal P}$ is not invariant under arbitrary local phase (gauge)  
transformations, i.e. $|\psi (t) \rangle \longrightarrow \exp (if({\bf Q},t)) 
|\psi (t) \rangle$, was one of the motivating factors \cite{wanelik} in the 
subsequent work of Aitchison and Wanelik 
\cite{aitchison}, who defined a phase associated with arbitrary, non-cyclic
evolutions and denoted here by $\gamma^{AW}$:
\begin{equation}
\exp (i\gamma^{AW} [\psi ]) = \left( \frac{\langle \psi (0) |\psi (T) 
\rangle}{\langle \psi (T) |\psi (0) \rangle} \right)^{1/2} \exp \left( 
-\int_{0}^{T}  \langle \psi(t) |\frac{d}{dt}|\psi(t) \rangle dt \right) ,
\label{eq:AW}
\end{equation}
where now the argument in the functional $\gamma^{AW} [\cdot ]$ is, in general,
a noncyclic path in ${\cal H}$.
We are assuming here as above that the states are normalised. The 
Aitchison-Wanelik phase factor (\ref{eq:AW}) is also geometric in the above 
sense (reparametrisation invariant and projective-geometric), and reduces 
to the Aharonov-Anandan phase factor (\ref{eq:AA}) in the case of 
cyclicity. Note that the Aitchison-Wanelik phase for an arbitrary 
open curve in ${\cal P}$ is actually numerically equal to
the Aharonov-Anandan phase obtained by geodesic closure of the 
curve\footnote{An earlier attempt to define a geometric 
phase for non-cyclic evolutions based on the idea of geodesic closure, was 
given by Samuel and Bhandari \cite{samuel}. However, as was pointed out in 
\cite{aitchison}, Samuel and Bhandari never departed from the 
Aharonov-Anandan phase since the geodesic closure makes the phases 
conceptually identical.}. 

The question now arises whether this phase, which is well-defined in all 
frames, is Galilean invariant. It is shown in the following that this
is not the case.

Consider the Galilean subgroup consisting of boosts in, say, the 
$x$-direction. That is, we consider two 
inertial frames, $S$ and $\tilde{S}$, associated 
with coordinate systems in the
standard configuration, the motion of $\tilde{S}$ relative to $S$  being    
of velocity ${\bf v}$ and parallel to the $x$-axis. In this case, it is 
straightforward to derive the following identities
\begin{eqnarray}
U_G^{\dagger}Q_iU_G & = & Q_i-vt\delta_{ix} \nonumber\\
U_G^{\dagger}P_iU_G & = & P_i-mv\delta_{ix} , 
\label{eq:optransf} 
\end{eqnarray}
where $i=x,y,z$, $\delta_{ix}$ is the Kronecker symbol and $v=|{\bf v}|$.

We are interested in the transformed Aitchison-Wanelik phase, i.e. geometric
phase for the ket $|\tilde{\psi}\rangle =U_{G}|\psi \rangle$: 
\begin{equation}
\exp (i\gamma^{AW} [\tilde{\psi}]) = \left( \frac{\langle \tilde{\psi} 
(0) |\tilde{\psi} (T) \rangle}
{\langle \tilde{\psi} (T) |\tilde{\psi} (0) \rangle} 
\right)^{1/2} \exp \left( 
-\int_{0}^{T} \langle \tilde{\psi}(\tilde{t}) |
\frac{d}{d\tilde{t}}|{\tilde{\psi}}(\tilde{t}) \rangle d\tilde{t} \right).
\end{equation}
Using the unitary operator $U_G$ in (\ref{eq:passive}) and the results 
(\ref{eq:optransf}) we obtain
\begin{eqnarray}
\exp (i\gamma^{AW}[\tilde{\psi}])& =& 
\left( \frac{\langle \psi(0)|U_G^{\dagger}(0) 
U_G(T)|\psi(T) \rangle}
{\langle \psi(T)|U_G^{\dagger}(T) U_G(0)|\psi(0) 
\rangle} \right)^{1/2} \exp \left( 
-\int_{0}^{T} \langle \psi (t)|U_G^{\dagger}(t)\frac{d}{dt}(
U_G(t)|\psi(t) 
\rangle )dt \right) \nonumber \\ 
& =&\left( \frac{\langle \psi (0) | e^{ivP_xT/\hbar} 
|\psi (T) \rangle} {\langle \psi (T) | e^{-ivP_xT/\hbar} 
|\psi (0) \rangle} 
\right)^{1/2}\exp \left(-\frac{imv^2T}{2\hbar}\right)\nonumber \\
& &\times \exp \left( -\int_0^T
\langle \psi(t)|U_G^{\dagger}(t) \frac{dU_G(t)}{dt}|\psi(t) \rangle dt 
\right) \exp \left( -\int_{0}^{T}\langle \psi (t) |\frac{d}{dt}| 
\psi (t)\rangle dt \right) \nonumber \\ 
& =&\left( \frac{\langle \psi (0) | e^{ivP_xT/\hbar} 
|\psi (T) \rangle} {\langle \psi (T) | e^{-ivP_xT/\hbar} 
|\psi (0) \rangle} \right)^{1/2}
\exp \left( -\frac{iv}{\hbar}\int_{0}^{T} \langle \psi (t)|P_{x}| \psi (t)
\rangle dt \right) \nonumber \\
 & & \times \exp \left( -\int_{0}^{T}\langle \psi (t)|\frac{d}{dt}| 
\psi (t)\rangle dt \right),
\label{eq:transfnoncyclic1}
\end{eqnarray}
where we have used $U_G^{\dagger}dU_G/dt=-imv^{2}/(2\hbar)+iP_{x}v/\hbar$
from (\ref{eq:passive}). If we compare (\ref{eq:AW}) and 
(\ref{eq:transfnoncyclic1}) we get 
\begin{eqnarray}
\exp (i\gamma^{AW}[\tilde{\psi}]) & = & \exp (i\gamma^{AW}[\psi ] ) 
\left( \frac{\langle \psi (0) | e^{ivP_xT/\hbar} 
|\psi (T) \rangle}{\langle \psi (0) |\psi (T) \rangle} 
\frac{\langle \psi (T) |\psi (0) \rangle}
{\langle \psi (T) | e^{-ivP_xT/\hbar} |\psi (0) \rangle} 
\right)^{1/2} \nonumber \\
 & & \times \exp \left( -\frac{iv}{\hbar}\int_{0}^{T} 
\langle \psi (t)|P_x| \psi (t)\rangle dt  \right) . 
\label{eq:transfnoncyclic2}
\end{eqnarray}
Now it is straightforward to show that 
\begin{equation}
\langle \psi |P_x| \psi \rangle = \langle \psi |A_{x}| \psi \rangle +
m \frac{d}{dt} \langle \psi |Q_x| \psi \rangle 
\label{eq:ehrenfest}
\end{equation}
where $A_{x} = A_{x} ({\bf Q},t)$ is the $x-$component of the vector potential
(if any) appearing in the Hamiltonian during the interval $[0,T]$.
So from (\ref{eq:transfnoncyclic2}) and (\ref{eq:ehrenfest}) we have
\begin{eqnarray}
\exp (i\gamma^{AW}[\tilde{\psi}]) & = & \exp (i\gamma^{AW}[\psi ] ) 
\left( \frac{\langle \psi (0) | e^{ivP_xT/\hbar} 
|\psi (T) \rangle}{\langle \psi (0) |\psi (T) \rangle} 
\frac{\langle \psi (T) |\psi (0) \rangle}
{\langle \psi (T) | e^{-ivP_xT/\hbar} |\psi (0) \rangle} 
\right)^{1/2} \nonumber \\
 & & \times \exp \left( -\frac{iv}{\hbar} \int_{0}^{T} \langle \psi (t)
|A_{x} ({\bf Q},t)| \psi (t) \rangle dt \right) \nonumber \\
 & & \times \exp \left( -\frac{imv}{\hbar} \left( \langle \psi (T)|Q_x| 
\psi (T) \rangle -\langle \psi (0)|Q_x| \psi (0) \rangle\right) \right) . 
\label{eq:transfnoncyclic3}
\end{eqnarray}
The last phase factor on the RHS of (\ref{eq:transfnoncyclic3}) is gauge 
independent, and will be unity if there exists one gauge such that cyclicity 
holds relative to $S$. The middle phase factor will clearly be unity when
$\int_{0}^{T} \langle \psi |A_{x}| \psi \rangle dt$ vanishes. (This happens
whenever, e.g., $A_{x}=0$ during $[0,T]$, and a gauge can always be chosen 
which ensures this condition.) 

Let us then finally consider the case where there exists a gauge such that
cyclicity holds relative to $S$, and that in the chosen gauge - which is
not necessarily this `cyclic' gauge - it transpires that $\int_{0}^{T}
\langle \psi |A_{x}|\psi \rangle dt$ vanishes. Then 
\begin{equation}
\exp (i\gamma^{AW} [\tilde{\psi}]) = \exp (i\gamma^{AW} [\psi ])
\left( \frac{\langle \psi (0)|e^{ivP_{x}T/\hbar}| \psi (T) \rangle}
{\langle \psi (0)|\psi (T) \rangle} \frac{\langle \psi (T)|\psi (0)\rangle}
{\langle \psi (T)|e^{-ivP_{x}T/\hbar} |\psi (0) \rangle} \right)^{1/2} .
\end{equation}
Given that $P_{x}$ is the generator of translations in the $x-$direction, 
it is evident here that the Galilean non-invariance of geometric phase is 
linked to the spatial displacement $vT$ at $t=T$ of the coordinate systems 
adapted to $S$ and $\tilde{S}$. The conceptual implications of this 
non-invariance, in particular in the context of measurements of geometric 
phase, will be dealt with elsewhere.
\vskip 0.5 cm
One of us (HRB) wishes to thank Joy Christian, Peter Holland, 
Kazimir Wanelik and most of all Jeeva Anandan for helpful discussions.
ES acknowledges post-doctoral scholarships from The Royal Swedish Academy of 
Sciences (Per Erik Lindahl's Foundation) and The Wenner-Gren Center 
Foundations.

\end{document}